\documentclass[aps,superscriptaddress]{revtex4}
\usepackage{amssymb,amsmath,epsfig}
\usepackage{subfigure}
\usepackage[colorlinks=true, pdfstartview=FitV, linkcolor=blue, citecolor=red, urlcolor=magenta, breaklinks=true]{hyperref}
\begin{document}
\title{Noncommutative correction to the entropy of Schwarzschild black hole with GUP}
\author{M. A. Anacleto}
\email{anacleto@df.ufcg.edu.br}
\affiliation{Departamento de F\'{\i}sica, Universidade Federal de Campina Grande
Caixa Postal 10071, 58429-900 Campina Grande, Para\'{\i}ba, Brazil}
\author{F. A. Brito}
\email{fabrito@df.ufcg.edu.br}
\affiliation{Departamento de F\'{\i}sica, Universidade Federal de Campina Grande
Caixa Postal 10071, 58429-900 Campina Grande, Para\'{\i}ba, Brazil}
\affiliation{Departamento de F\'isica, Universidade Federal da Para\'iba, Caixa Postal 5008, 58051-970 Jo\~ao Pessoa, Para\'iba, Brazil}
\author{S. S. Cruz}
\email{sergiomusiarte@hotmail.com}
\affiliation{Departamento de F\'{\i}sica, Universidade Federal de Campina Grande
Caixa Postal 10071, 58429-900 Campina Grande, Para\'{\i}ba, Brazil}
\author{E. Passos}
\email{passos@df.ufcg.edu.br}
\affiliation{Departamento de F\'{\i}sica, Universidade Federal de Campina Grande
Caixa Postal 10071, 58429-900 Campina Grande, Para\'{\i}ba, Brazil}

\begin{abstract} 
In this paper we study through tunneling formalism, the effect of noncommutativity to Hawking radiation and the entropy of the noncommutative Schwarzschild black hole. 
In our model we have considered the noncommutativity implemented via the Lorentzian distribution. 
We obtain noncommutative corrections to the Hawking temperature using the Hamilton-Jacobi method 
and the Wentzel-Kramers-Brillouin (WKB) approximation.
In addition, we found corrections of the logarithmic and  other types due to noncommutativity and quantum corrections from the generalized uncertainty principle (GUP) for the entropy of the Schwarzschild black hole.

\end{abstract}
\maketitle
\pretolerance10000

\section{Introduction}
In the study of quantum gravity (a merger of general relativity and quantum mechanics) it may reveal new features of the black hole near the Planck scale.
An interesting prediction of quantum gravity theories such as loop quantum gravity, string theory and noncommutative geometry, as well as black hole physics corresponds to the existence of a minimum length~\cite{Garay:1994en,AmelinoCamelia:2000ge,Ali:2009zq}.
In this way the effect of the minimum length must also be considered in the uncertainty principle and as a result of this we will have the so-called generalized uncertainty principle (GUP). 
In recent years, the effect of GUP on the thermodynamics of various types of black holes has been extensively analyzed~\cite{Yang:2018wlb,Anacleto:2014apa,Feng:2015jlj,Anacleto:2019rfn,Haldar:2019fcz,Javed:2019btd,Li:2016mwq,Gecim:2017zid,Chen:2016ftz,Maluf:2018lyu,Gomes:2018oyd}. 
Consequently, it has been found in many cases that the GUP has an important role in black hole physics, which is the emergence of black hole remnants~\cite{Adler:2001vs,Chen:2002tu,Dutta:2014yna,Anacleto:2015rlz,Chen:2014jwq,Mandal:2018xom,Anacleto:2015kca,Kuntz:2019gka,Li:2017zbc}.
{In addition, in determining the entropy of black holes it has been found logarithmic corrections to the area law due to the effect of quantum gravity}~\cite{Anacleto:2015mma,Anacleto:2015awa,El-Menoufi:2015cqw,Jusufi:2016nxo,Sadeghi:2016xym,Giardino:2020myz}.
In~\cite{Li:2020gzi}, by considering the effect of the Lorentz invariance violation of dispersion relations, the Hawking radiation of the charged fermions via tunneling from the Reissner-Nordstr\"{o}m black hole has been investigated.

On the other hand, there are several ways in the literature on how to implement a noncommutative spacetime in theories of gravity~\cite{Nicolini:2008aj,Aschieri:2005yw,Aschieri:2005zs,Meljanac:2007xb,Meljanac:2006ui,Harikumar:2012zi}. In particular, noncommutativity has been introduced in black hole physics through a modification of the matter source by a Gaussian mass distribution~\cite{Nicolini:2005vd} or by a Lorentzian mass distribution~\cite{Nozari:2008rc}.
However, applications in noncommutative black hole physics implemented via Lorentzian distribution have been very little explored. In~\cite{Kim:2008vi} the thermodynamic similarity between the noncommutative Schwarzschild black hole and the Reissner-Nordstr\"{o}m black hole has been analyzed. 
By adopting the tunneling formalism the thermodynamics of noncommutative black holes {have been investigated} in~\cite{Nozari:2009nr,Banerjee:2008gc,Mehdipour:2009zz,Mehdipour:2010kp,Mehdipour:2010ap,Miao:2010wy,Miao:2011dy,Nozari:2012bp,Gupta:2013ata,Ovgun:2015box,Gupta:2017lwk,Gecim:2020zcb}. 
In~\cite{Liang:2012vx} by taking the mass density to be a Lorentzian smeared
mass distribution the thermodynamic properties of noncommutative BTZ black holes {have been studied} 
and in~\cite{Anacleto:2019tdj,Anacleto:2013-analogue} the process of massless scalar wave scattering by a noncommutative black hole was examined. 
In~\cite{prep}, logarithmic corrections have been shown to appear due to the noncommutative and GUP effect for the BTZ black hole.

In present paper, inspired by all of these previous works, we examine the thermodynamics quantities of a noncommutative Schwarzschild black hole with GUP using the Hamilton-Jacobi method and the WKB approximation in the tunneling formalism.
In this way we implemented noncommutativity in the Schwarzschild black hole metric by adopting a mass density as being a Lorentzian smeared mass distribution.
Then we analyze the changes in thermodynamic properties due to the effects of noncommutativity and GUP for temperature, entropy and specific heat.
In addition, we verified the possibility of the existence of remnants in this set up. 
As a result of these calculations we found the presence of Schwarzschild black hole remnants in the noncommutative background with GUP and also in the absence of GUP.
Moreover, we also obtain noncommutative logarithmic corrections for entropy in the presence of the GUP and without the GUP.

The paper is organized as follows.
In Sec.~\ref{sec2} we incorporate the effect of noncommutativity in the Schwarzschild black hole metric via Lorentzian smeared mass distribution and we study the modification of the thermodynamic quantities.
In Sec.~\ref{sec3} using the GUP we determine the quantum corrections for the Hawking temperature, entropy and specific heat. 
In Sec.~\ref{conc} we make our final considerations.

\section{Noncommutative corrections to the Schwarzschild black holes}\label{sec2}
Our starting point here is to incorporate the contribution of noncommutativity into the Schwarzschild black hole metric by assuming a Lorentzian smeared mass distribution~\cite{Nicolini:2005vd,Nozari:2008rc,Anacleto:2019tdj}
as follows: 
\begin{eqnarray}
\varrho_{\theta}(r)=\frac{M\sqrt{\theta}}{\pi^{3/2}(r^2+\theta)^{2}},
\end{eqnarray}
where $ \theta $ is the noncommutative parameter whose dimension is given in terms of ${length}^2$ and $ M $ is the total mass diffused throughout a region
with linear size $ \sqrt{\theta} $. 
We can now determine the smeared mass distribution function through the relationship~\cite{Anacleto:2019tdj} 
\begin{eqnarray}
{\cal M}_{\theta}&=&\int_0^r\varrho_{\theta}(r)4\pi r^2 dr,
\\
&=&\frac{2M}{\pi}\left[\tan^{-1}\left( \frac{r}{\sqrt{\pi\theta}} \right)
-\frac{r\sqrt{\pi\theta}}{\pi\theta + r^2}  \right],
\\
&=&M-\frac{4 M\sqrt{\theta}}{\sqrt{\pi}r} + {\cal O}(\theta^{3/2}). 
\end{eqnarray}
In this case, the noncommutative Schwarzschild black hole metric is constructed by considering the modified mass above and thus we obtain
\begin{eqnarray}
ds^{2} = -{\cal N}(r)dt^{2} + {\cal N}(r)^{-1}dr^{2}+r^{2}d\phi^2,
\end{eqnarray}
with
\begin{eqnarray}
&&{\cal N}(r)=1-\frac{2{\cal M}_{\theta}}{r} 
= 1-\frac{2 M}{r} +\frac{8 M \sqrt{\theta}}{\sqrt{\pi}r^2}+{\cal O}(\theta^{3/2}).
\end{eqnarray}
The horizons are,  up to first order in $ \sqrt{\theta} $, given by
\begin{eqnarray}
\label{rh}
&&\tilde{r}_h=r_h-4\sqrt{\frac{\theta}{\pi}} + \cdots
=r_h\left(1-\frac{4}{r_h}\sqrt{\frac{\theta}{\pi}}\right), \quad r_h=2M,
\\
&&r_{\theta}=4\sqrt{\frac{\theta}{\pi}} + \cdots.
\end{eqnarray}
From equation (\ref{rh}) we can express the mass in terms of $ \tilde{r}_h $ as follows
\begin{eqnarray}
M=\frac{\tilde{r}_h}{2} + 2\sqrt{\frac{\theta}{\pi}}.
\end{eqnarray}
In order to determine the Hawking temperature and entropy we consider the Klein-Gordon equation 
for a scalar field $ \Phi $ in the curved space given by
\begin{eqnarray}
\left[\frac{1}{\sqrt{-g}}\partial _{\mu}(\sqrt{-g}g^{\mu \nu}\partial _{\nu}) 
- \frac{m^{2}}{\hbar ^{2}}\right] \Phi = 0 ,
\label{EqKG}
\end{eqnarray}
that using the WKB approximation
\begin{eqnarray}
\Phi = \exp\left[\frac{i}{\hbar}{\cal I}(t,r,x^{i})\right],
\label{WKB}
\end{eqnarray}
we can find
\begin{eqnarray}
-\frac{1}{{\cal N}(r)}(\partial _{t}{\cal I})^{2} + {\cal N}(r)(\partial _{r}{\cal I})^{2} 
+ \frac{1}{r^{2}}(\partial _{\phi}{\cal I})^{2} + m^{2} = 0.
\label{KG-WKB}
\end{eqnarray}
Now we can make a separation of variables into the equation (\ref{KG-WKB}) as follows
\begin{eqnarray}
{\cal I} = -{\cal E}t + {\cal R}(r) + J_{\phi}\phi,
\label{I}
\end{eqnarray}
being
\begin{eqnarray}
\partial_t {\cal I}=-{\cal E}, \quad \partial_r {\cal I}=\frac{d{\cal R}(r)}{dr}, \quad 
\partial_{\phi}{\cal I}=J_{\phi}.
\end{eqnarray}	
So as a result we find
\begin{eqnarray}
{\cal I} = -{\cal E}t + \int dr \frac{\sqrt{{\cal E}^2 - {\cal N}(r)\left(\frac{J_{\phi}^2}{r^2} + m^2\right)}}{{\cal N}(r)} + J_{\phi}\phi.
\label{I2}
\end{eqnarray}
Then, to determine $ {\cal R}(r) $ we consider in the near event horizon regime, $ r\rightarrow \tilde{r}_{h} $, the following approximation $ {\cal N}(r)\approx\kappa (r - \tilde{r}_{h}) $, so we have
\begin{eqnarray}
	{\cal R}(r) =\frac{1}{\kappa}\int dr \frac{\sqrt{{\cal E}^2 - \kappa(r - \tilde{r}_{h}) 
	\left(\frac{J_{\phi}^2}{r^2} 
	+ m^2\right)}}{(r-\tilde{r}_{h})}
	=\frac{2\pi i}{\kappa}{\cal E},
	\label{W}
\end{eqnarray}
being $ \kappa $ the surface gravity of the noncommutative Schwarzschild black hole given by
\begin{eqnarray}
\kappa = {\cal N}'(\tilde{r}_{h}) = \frac{r_{h}}{\tilde{r}^2_h}
-\frac{8r_h\sqrt{\theta}}{\sqrt{\pi}\tilde{r}^3_h}.
\label{sgrav}
\end{eqnarray}
For the tunneling probability we can find
\begin{eqnarray}
\Gamma \simeq \exp[-2{\rm\, Im}({\cal I})]=\exp\left(-\frac{4\pi {\cal E}}{\kappa}\right),
\label{Gamma}
\end{eqnarray}
that comparing (\ref{Gamma}) with the Boltzmann factor  $\exp({-{\cal E}/{\tilde{T}_{H}}})$ we obtain
\begin{eqnarray}
\label{ThB}
{\cal T}_{H}&=& \frac{\kappa}{4\pi}
=\frac{1}{4\pi}\left(\frac{r_{h}}{\tilde{r}^2_h}
-\frac{8r_h\sqrt{\theta}}{\sqrt{\pi}\tilde{r}^3_h}\right),
\\
&=&\frac{1}{4\pi\tilde{r}_h}-\frac{\sqrt{\theta}}{\pi\sqrt{\pi}\tilde{r}^2_h}
-\frac{8\theta}{{\pi}^2\tilde{r}^3_h}.
\label{TH}
\end{eqnarray}
For $ \theta=0 $ the temperature of the Schwarzschild black hole, $ T_H=1/4\pi r_h $ is recovered.

Next, from the first law of black hole thermodynamics we proceed to compute the entropy that reads
\begin{eqnarray}
{\cal S}=\int \frac{dM}{{\cal T}_H}=\int\frac{\partial M}{\partial\tilde{r}_h} 
\frac{d\tilde{r}_h}{{\cal T}_H},
\end{eqnarray}
where
\begin{eqnarray}
{\cal T}^{-1}_H=4\pi\tilde{r}_h\left(1+\sqrt{\frac{\theta}{\pi}}
\frac{4}{\tilde{r}_h}+\frac{32\theta}{\pi\tilde{r}^2_h}+\cdots \right), 
\quad \frac{\partial M}{\partial\tilde{r}_h}=\frac{1}{2}.
\end{eqnarray}
Thus, the entropy of the noncommutative black hole gets the following corrections:
\begin{eqnarray}
{\cal S}&=&\int\frac{1}{2}\left(4\pi\tilde{r}_h + 16\sqrt{\pi\theta} 
+ \frac{128\theta}{\tilde{r}_h}+\cdots\right) d\tilde{r}_h,
\\
&=& \frac{{\cal A}}{4} + 8\sqrt{\pi\theta}\tilde{r}_h + 64\theta\ln(\tilde{r}_h) +\cdots,
\\
&=&\frac{{\cal A}}{4} + 4\sqrt{\theta {\cal A}} + 32\theta\ln\left(\frac{{\cal A}}{4\pi}\right)+\cdots,
\end{eqnarray}
where ${\cal A}=4\pi\tilde{r}^2_h$ is the noncommutative Schwarzschild black hole horizon area. 
Note that noncommutative corrections of the logarithmic, $ \theta\ln{\cal A} $ and $ \sqrt{\theta {\cal A}} $ type are obtained for entropy. 
When $ \theta=0 $ we recover the result for the entropy of the Schwarzschild black hole.

\section{Entropy with GUP}\label{sec3}
At this point in order to compute the thermodynamic quantities, Hawking temperature, entropy and specific heat of the noncommutative Schwarzschild black hole we will use the GUP~\cite{ADV, Tawfik:2014zca, Dutta:2015, KMM, bastos} given by
\begin{eqnarray}
\label{gup}
\Delta x\Delta p\geq \frac{\hbar}{2}\left( 1-\frac{\alpha l_p}{\hbar} \Delta p +\frac{\alpha^2 l^2_p}{\hbar^2} (\Delta p)^2 \right),
\end{eqnarray}
being $ l_p $ the Planck length and $\alpha$ a dimensionless positive parameter .

Now for the purpose of obtaining a dispersion relation corrected by the GUP for the energy of the black hole, we assume that $ \Delta p\sim {\cal E} $  and without loss of generality we adopt the natural units system,  i.e.,
$ G=c=k_B=\hbar=l_p=1 $. Therefore, applying the same steps performed in~\cite{Anacleto:2019rfn} we have
\begin{eqnarray}
{\cal E}_{gup}\geq {\cal E}\left[1-\frac{\alpha}{2(\Delta x)}+ \frac{\alpha^2}{2(\Delta x)^2}+\cdots    \right].
\end{eqnarray}
In sequence, to find the tunneling probability we follow the same steps performed in the previous section and thus obtain
\begin{eqnarray}
\Gamma\simeq \exp[-2{\rm Im} ({\cal I})]=\exp\left[\frac{-4{\pi}{\cal E}_{gup}}{a}\right],
\end{eqnarray}
being $ a $ the surface gravity. 
Finally, by comparing with the Boltzmann factor, the modified Hawking temperature of the noncommutative Schwarzschild black hole can be recast in the following form
\begin{eqnarray}
{\cal T}_{gup}\leq {\cal T}_H\left[ 1-\frac{\alpha}{2(\Delta x)}+ \frac{\alpha^2}{2(\Delta x)^2}+\cdots   \right]^{-1}.
\end{eqnarray}
Moreover, assuming that in the vicinity of the event horizon the minimum uncertainty is of the order of the horizon radius, the Hawking temperature modified by the GUP can be rewritten as
\begin{eqnarray}
\label{Tgup}
{\cal T}_{gup}&\leq& {\cal T}_H\left(1 - \frac{\alpha}{4\tilde{r}_{h}} + \frac{\alpha ^{2}}{8\tilde{r}_{h}^{2}}+\cdots\right)^{-1},
\\
&=&\left(\frac{1}{4\pi\tilde{r}_h}-\frac{\sqrt{\theta}}{\pi\sqrt{\pi}\tilde{r}^2_h}
-\frac{8\theta}{{\pi}^2\tilde{r}^3_h}  \right)
\left(1 + \frac{\alpha}{4\tilde{r}_{h}} - \frac{\alpha ^{2}}{8\tilde{r}_{h}^{2}}+\cdots\right).
\end{eqnarray}
Next, we will compute the entropy of the noncommutative Schwarzschild black hole by using the following formula:
\begin{eqnarray}
S_{gup} &=&\int \frac{1}{{\cal T}_{gup}}\frac{\partial M}{\partial\tilde{r}_h}d\tilde{r}_{h}.
\end{eqnarray}
So, as a result we obtain
\begin{eqnarray}
S_{gup} &=&\int\frac{1}{2}\left(4\pi\tilde{r}_h + 16\sqrt{\pi\theta} 
+ \frac{128\theta}{\tilde{r}_h}+\cdots\right)
\left[1 - \frac{\alpha}{4\tilde{r}_{h}} + \frac{\alpha ^{2}}{8\tilde{r}_{h}^{2}}+\cdots\right]d\tilde{r}_{h},
\\
&=&\frac{{\cal A}}{4} + 8\sqrt{\pi\theta}\tilde{r}_h + 64\theta\ln(\tilde{r}_h) -\frac{\pi\alpha\tilde{r}_h}{2}
+\frac{\pi\alpha^2}{4}\ln(\tilde{r}_h)
\nonumber\\
&-&2\sqrt{\pi\theta}\alpha\ln(\tilde{r}_h) - \frac{\sqrt{\pi\theta}\alpha^2}{\tilde{r}_h}
+\frac{16\theta\alpha}{\tilde{r}_h}- \frac{4\theta\alpha^2}{\tilde{r}_h^2}+\cdots.
\end{eqnarray}
Besides, using the Hamilton-Jacobi method in tunneling formalism, we have derived the modified Hawking temperature and noncommutative/quantum corrections for the Schwarzschild black hole entropy. 
In addition, in the result above we see that logarithmic corrections and other types were obtained for the area law.
On the other hand, in the absence of GUP, i.e., for $\alpha =0 $, noncommutative corrections for entropy can be presented as follows
\begin{eqnarray}
{\cal S}_{\theta}
&=&\frac{{\cal A}}{4} + 8\sqrt{\pi\theta}\tilde{r}_h + 64\theta\ln(\tilde{r}_h) +\cdots.
\end{eqnarray}
Now we can express the result above in terms of $ r_h $ as follows
\begin{eqnarray}
{\cal S}_{\theta}
&=&\frac{A}{4} -16\theta + 64\theta\ln({r}_h) +\cdots,
\\
&=&\frac{A}{4} -16\theta + 32\theta\ln(A/4\pi) +\cdots.
\end{eqnarray}
Here $ A=4\pi r^2_h $ is identified as the Schwarzschild black hole area.
Hence, we obtain a logarithmic correction for entropy due to the effect of the noncommutative background.
Furthermore, we see that noncommutative corrections for Schwarzschild black hole entropy occur  at the first order of  $\theta $.
{Moreover, it is worth mentioning that subleading logarithmic corrections also appear in the calculation of entanglement entropy~\cite{Solodukhin:2011gn}. As in entanglement entropy, the logarithm terms here have a similar UV origin, since one expects that noncommutativity and quantum corrections from the generalized
uncertainty principle (GUP) may be probed only in very short distances (or high energy scales).
Furthermore, as argued in~\cite{Solodukhin:2011gn}, the logarithm term is important in investigating the evaporation process of the final stage of the black hole. Thus, the black hole reaches a minimum mass in the final evaporation stage. Below, we find the minimum mass of the noncommutative black hole through the specific heat capacity.}

Now for $\alpha=0 $, the noncommutative correction for specific heat capacity is given by
\begin{eqnarray}
\label{shc}
{\cal C}&=&\frac{\partial M}{\partial {\cal T}_H}
=\frac{\partial M}{\partial\tilde{r}_h}\left(\frac{\partial {\cal T}_H}{\partial\tilde{r}_h}  \right)^{-1},
\\
&=&-2\pi\tilde{r}^2_h-16\sqrt{\pi\theta}\tilde{r}_h-24\theta+\cdots,
\end{eqnarray}
that presenting in terms of $ r_h $  we have 
\begin{eqnarray}
\label{Cvtheta}
{\cal C}&=& -2\pi\left( {r}^2_h - \frac{4\theta}{\pi}\right).
\end{eqnarray}
It is worth mentioning that, from the expression above, we can obtain the condition for the formation of a black hole remnant consisting of $ {\cal C} = 0 $ where the evaporation process stops.
So taking $r_{h} ={2\sqrt{\theta/\pi}}$ the specific heat is zero.
Further, this implies a minimum radius 
\begin{eqnarray}
r_{min}=2M_{min}={2\sqrt{\theta/\pi}}.
\end{eqnarray}
Consequently, we obtain the following minimum mass for the noncommutative Schwarzschild black hole:
\begin{eqnarray}
M_{min}=\sqrt{\frac{\theta}{\pi}}. 
\end{eqnarray}
Now, by substituting the $ r_ {min} $ in (\ref{ThB}) we obtain a maximum temperature given by
\begin{eqnarray}
{\cal T}_{max}\approx 5T_H=\frac{5}{4\pi r_{min}}=\frac{5}{8\sqrt{\pi\theta}}.
\end{eqnarray}
Therefore, we show that the effect of noncommutativity implies the existence of a black hole remnant.
In this case, the presence of the minimum mass (or minimum radius) shows that the specific heat goes to zero and the black hole stops evaporating even reaching a maximum temperature.

\section{CONCLUSIONS}\label{conc}
In this paper we have examined the effect of noncommutativity and GUP on the 
noncommutative Schwarzschild black hole thermodynamics through tunneling formalism.
We have incorporated the contribution of noncommutativity to the noncommutative Schwarzschild black hole metric by means of a Lorentzian  mass distribution.
By computing the entropy, we have found a logarithmic corrections due to both the contribution of noncommutativity and quantum corrections from the GUP.  This signalizes the well-known corrections to the entropy given by $S_{gup}\sim A/4 + (c_1 + c_2) \ln A + ...,$ where the `species' $c_i$ here are essentially related to each parameter of correction corresponding to noncommutativity and GUP. Moreover, we have verified the existence of black hole remnants by calculating the specific heat. The results show  that even in the absence of GUP the black hole becomes a remnant reaching a maximum temperature.

\acknowledgments
We would like to thank CNPq, CAPES and CNPq/PRONEX/FAPESQ-PB (Grant nos. 165/2018 and 015/2019),  for partial financial support. MAA, FAB and EP acknowledge support from CNPq (Grant nos. 306962/2018-7 and  433980/2018-4, 312104/2018-9, 304852/2017-1).

\end{document}